\documentclass[
aps,pre,reprint, superscriptaddress, floatfix, 
dvipdfmx
]{revtex4-2}

\usepackage{amsthm}
\usepackage{amsmath} 
\usepackage{amssymb}
\usepackage{physics}
\usepackage[utf8]{inputenc}
\usepackage[dvipdfmx]{graphicx}
\usepackage{dcolumn}
\usepackage{bm}
\usepackage{dsfont}
\usepackage[german, greek, french, english]{babel}
\usepackage[dvipsnames]{xcolor}
\usepackage{mathtools}
\usepackage{comment}
\usepackage{tabularx}
\usepackage{lipsum}
 \usepackage{booktabs}
 
 \usepackage[dvipdfmx]{hyperref}
 \hypersetup{
colorlinks=true,
urlcolor= blue,
citecolor=blue,
linkcolor= blue,
}

\addto\captionsenglish{}
\addto\captionsenglish{}

\newcommand{\ens}[1]{\left< {#1} \right>}

\DeclareMathOperator{\sgn}{sgn}
 \newcolumntype{Y}{>{\centering\arraybackslash}X}
 
\begin{document}

\title{
Modeling urbanization dynamics by labor force migration
}

\author{Hirotaka Goto}
\email[]{hirotakagoto@meiji.ac.jp}
\affiliation{
 Graduate School of Advanced Mathematical Sciences, Meiji University, 
 4-21-1 Nakano, Nakano City, Tokyo 164-8525, Japan}

\date{October 23, 2024}

\begin{abstract}

Individual participants in human society collectively exhibit aggregation behavior. 
In this study, we present a simple microscopic model of labor force migration based on the active Brownian particles framework.
In particular, agent-based simulations show that the model produces clusters of agents from a random initial distribution. 
Furthermore, two empirical regularities called Zipf's and Okun's laws were observed. 
To reveal the mechanism underlying the reproduced aggregation phenomena, 
we use our microscopic model to derive an extended Keller--Segel system, 
which is 
a classic model describing the aggregation behavior of biological organisms called \emph{taxis}. 
The obtained macroscopic system indicates that the concentration of the workforce in the real world can be explained 
through a new type of taxis central to human behavior, 
highlighting the relevance of urbanization to blow-up phenomena in the derived PDE system. 
We then characterize the transition between the aggregation and diffusion regimes both analytically and computationally. 
The predicted long-term dynamics of urbanization---originating in the asymmetric natures of 
employed and unemployed agents---are compared 
with global empirical data, 
particularly in the realms of labor statistics and urban indicators.

\end{abstract}

\maketitle

\section{\label{sec: introduction} Introduction}

Urbanization---a process of population concentration \cite{Tisdale1942}---is 
recognized as a significant characteristic of human society \cite{Berry2008, Boddy1999, Ottaviano2004}. 
For as long as human activities have been recorded, there has been a persistent trend toward urbanization, 
with a notable acceleration in the last few centuries \cite{Tisdale1942, Fujita2003, Davis1955}. 
The topic has been explored 
in numerous fields, including geography, economics, and urban planning, from various perspectives. 
Nevertheless, how people get concentrated might be widely universal, provided that similar agglomeration patterns
and hierarchical structures have been discovered across the globe \cite{Fujita1996, Brakman2003, Berry2008}.

In the last two centuries, various theories have been proposed 
to explain the spatial distributions of human activity and settlements, 
such as von Th{\"u}nen's theory \cite{Grotewold1959, Maki2004}, 
Weberian location theory \cite{Isard1949, Capello2021}, and the central place theory \cite{King1985, Mulligan1984}. 
Toward the end of the last century, in economics, a new discipline called New Economic Geography 
was established, and spatial economic problems, such as spatial distribution of industrial clusters, 
began to be considered seriously \cite{Ottaviano2004, Martin1999}. 
Currently, the nonuniform distribution of economy-related activity is most commonly understood as 
either a cost-minimization (or profit-maximization) problem or an equilibrium state, where the benefits
of agglomeration and the resulting costs balance each other 
\cite{Glaeser2010, Puga2010, Duranton2004, Boddy1999, Ottaviano2004}.

However, theories proposed in previous studies 
do not necessarily provide a clear understanding of the spatial concentration of populations 
\cite{Berry2008, Fujita1996}. 
For example, they have been criticized for their unestablished empirical basis, 
unclear microscopic underpinnings, lack of circular causation, little reference to intraregional dynamics, and lack of 
emergent nature of urban areas \cite{King1985, Tisdale1942, Martin1999, Boddy1999, Krugman1992}. 
Furthermore, theoretical arguments derived from these theories are not always verifiable \cite{Duranton2004, Martin1999} 
because they have often been formulated in an oversimplified manner or under implausible assumptions, 
thereby making a comparison with empirical data unfeasible \cite{Scott2004}. 
Moreover, these theories often tend to be one-sided in a sense 
that some primarily focus on the economic aspects of the phenomena 
while others make light of such perspectives in exchange for a disproportionate amount of interest 
in regional aspects of the problem. 
However, because urbanization encompasses nearly all facets of human activity, 
not only geographical and economic but cognitive and psychological perspectives 
should be present in any good model of urbanization.  
Integrating wide-ranging evidence and multiple sides of the phenomena in a coherent manner is critically important for 
understanding the emergence of such complex phenomena as urbanization. 
Although the task is not self-explanatory, it should most likely be achieved 
by constructing a foundational mathematical model in line with empirical observations.

Over the past few decades, statistical physics has proven to be a powerful tool 
for examining and understanding social dynamics \cite{Castellano2009}. 
A central example is human mobility patterns 
\cite{Gonzalez2008, Starnini2013, Shida2020, Schlapfer2021}. 
The study of human mobility using statistical physics 
has recently turned into an emerging field due in part to the increasing availability of large-scale data, 
e.g., location data recorded in mobile phones or some wearable devices. 
However, most studies in this field focus on short-term migration patterns. 
The reason is that studying human migration spanning a few centuries, which is a cause of urbanization, 
entails a significant amount of data that limits the development of such data-driven research. 
Therefore, a theoretical foundation is required for investigating several-centuries-long human mobility using statistical physics.

\begin{table*}[htbp] 
    \caption{List of models and theories explaining the uneven spatial distributions of populations and/or activities.}
    \label{tab: previous models}
    \begin{tabularx}{2\columnwidth}{l*{1}{Y}} 
	\hline \hline
	Theoretical framework & Key components \\ 
	\hline  
	Von Th{\"u}nen model \cite{Grotewold1959, Maki2004} 
	& Minimizing transportation costs \\
	Weberian location theory \cite{Isard1949, Capello2021} & Minimizing total costs (including transportation) \\
	Central place theory \cite{King1985, Mulligan1984} 
	& Hierarchically ordered urban places \\ 
	Spatial economics approach \cite{Ottaviano2004, Martin1999} 
	& Increasing returns to scale and transportation costs \\
	Law of motion of economy \cite{Krugman1992} 
	& Local wage (fitness) and local labor force growth \\
	Schweitzer's model \cite{Schweitzer1998} & Wage-oriented agent-based model of labor mobility \\
	Presently proposed model & Stimulus-driven agent-based model of labor mobility \\ 
	\hline \hline
    \end{tabularx}
\end{table*}

Consistent with this line of research, 
the author in \cite{Schweitzer1998} prominently developed 
a dynamic model of economic agglomeration 
by employing active Brownian particles \cite{Schweitzer2007, Romanczuk2012}, 
which were first conceptualized 
in \cite{SchimanskyGeier1995}. 
Active Brownian particles (sometimes referred to as Brownian agents \cite{Helbing2001}) 
are Brownian particles capable of generating a field that, 
in turn, influences their motion 
\cite{Romanczuk2012}. 
The study was also a generalization of another simple dynamic model of local economies \cite{Krugman1992}, 
previously proposed as part of an initial effort 
to bring to light the spatial aspects of economic activity, which had rarely been explored at the time in economics. 
In particular, the author in \cite{Schweitzer1998} generalized the model 
by explicitly considering the spatial movements of each individual agent, 
whereas in \cite{Krugman1992}, only local economic growth and decline were considered 
(see Table~\ref{tab: previous models} for a summary of previous models).

Using the active Brownian particle framework, 
the author in \cite{Schweitzer1998} investigated the emergence and evolution of ``economic centers'' 
and observed their coexistence. 
However, a major drawback is that the reproduced emergence of central regions is solely attributable to the 
strong nonlinearity of a modified production function, coupled with a rapid nonlinear 
response of the local employment status to changes in local economic circumstances. 
In other words, the model cannot reproduce aggregation phenomena 
unless all those specific nonlinear traits are satisfied, 
which might constrain the model substantially. 
It would also be less compatible with the idea of self-organization \cite{Camazine2001}, 
which was first mentioned in \cite{Krugman1992} from an economics perspective. 
In the present study, 
we will eliminate this issue mainly by redefining the driving force of the system in a simpler way that also aligns with 
a growing body of research in psychology and cognitive and neural science. 
More specifically, we will use the simple Cobb--Douglas production function \cite{Douglas1976}, 
which determines economic output, for example, as a function of labor, and Fechner's law \cite{Nutter2010}, 
which characterizes the typical response of individuals to a stimulus perceived. 
The limitation in the previous study will be discussed further in Sec.~\ref{subsec: previous model}, 
and the motivation behind the present model will be discussed in Sec.~\ref{sec: discussion}.

As with the tendency toward urbanization, 
society is endowed with other characteristic features, many of which have been
presented in the form of statistical regularities. 
The rank-size rule is a power-law relationship between the city size and its rank in a given urban system 
\cite{Hinloopen2006, Fujita1996, Rose2005}, described in the following mathematical form: 
\begin{align}
(\text{city size}) \propto (\text{rank})^{ - \gamma}, \label{eq: Zipf}
\end{align}
where the rank-size exponent $\gamma$ is close to 1 
\cite{Gabaix2004, Veneri2016}. 
This provides a good rule of thumb that the $N$th largest city has a population proportional to one-$N$th 
the population of the largest city in a community, which is also known as a good approximation of \emph{Zipf's law} 
\cite{Reed2002, Gabaix1999}. 
Zipf's law holds only for large cities \cite{Hinloopen2006}. 
This well-documented regularity is generally considered a manifestation of universality in human migration and settlements. 
In particular, several models have been proposed to replicate the law, such as Steindl's and Simon's models
\cite{Simon1955, Gabaix1999}. 
However, they have been criticized for their counterfactual settings or limited applicability, 
such as the longer time period required when seeking to obtain an exponent closer to $1$ 
\cite{Gabaix1999, Krugman1996}. 
A recent, more plausible explanation has been proposed by Gabaix \cite{Gabaix1999}, 
which connects Zipf's law to a simple underlying mechanism for city growth called \emph{Gibrat's law} \cite{Eeckhout2004}. 
The law posits stochastic city growth where the growth rates are determined by independent and identically distributed 
random variables with constant mean and variance. 
The detailed analysis is also given, for example, in \cite{Sornette2006}.

In relation to urbanization, empirical studies have indicated that a rank-size distribution demonstrating Zipf's law 
is a sign of urban development \cite{Cristelli2012, Wang2021, GuerinPace1995}, 
Several studies have reported a gradual but continuing increase in rank-size exponents (even after they reach $1$), 
such as the city-size distribution in China from 1982 to 2010 \cite{Wang2021} 
and the country-size distribution of the 50 largest countries worldwide from 1990 to 2050 (projections included) 
\cite{Rose2005}. 
A list of studies examining the time-evolution of rank-size exponents on a national level can be found in \cite{Delgado2004}.
This regularity appears on extensive and dynamic scales, thus requiring a more comprehensive theory of the law 
\cite{Rose2005}. 
Indeed, a recent meta-analysis \cite{Cottineau2017} has indicated that the evolution of city size distributions 
is more closely related to the urbanization process rather than economic or demographic development. 
The study then emphasized the necessity of generative models accounting for spatial patterns of migration, 
instead of non-spatial stochastic city growth models such as Gibrat's law.

\emph{Okun's law} is another statistical characteristic studied in macroeconomics literature \cite{Okun1963}. 
This law refers to the negative correlation between output growth and changes in the unemployment rate 
during an interval (e.g., quarterly) \cite{Lee2000}. 
That is, positive output growth corresponds to a decreasing unemployment rate. 
For empirical testing, the law can be expressed as follows \cite{Knotek2007}:
\begin{align}
\frac{\Delta Y_{tot}}{Y_{tot}} 
\approx - c \, \Delta \mu , \label{eq: Okun}
\end{align} 
where $Y_{tot}$ is the total output nationwide, 
and $\mu$ is the (national) average unemployment rate at a given time. 
$\Delta Y_{tot}$ and $ \Delta \mu$ represent changes 
in the respective variables during a given interval. 
In the case of the U.S. economy, 
approximately a 2\% increase in output is predicted (i.e., real GDP, empirically) 
for every one-point decrease in the unemployment rate, which translates to $c \approx 2$ 
\cite{Lee2000, Cuaresma2003}. 
An accumulated body of empirical and theoretical research 
has demonstrated that the law, including the coefficient, appears robust to some extent over different periods 
and under different methods of analysis \cite{Ball2017}. 
Although some variability exists in the coefficient among countries and controversy regarding the reliability of the law, 
the relationship is typical in most economies \cite{Lee2000}. 
Various existing theories consider fluctuations from hypothetical baseline values of certain 
variables not directly observable. 
However, these theories fail to 
account for empirically observed coefficients (e.g., the overestimation of the coefficients) \cite{Ball2017}. 
Therefore, a new, unified theoretical approach will be beneficial \cite{Knotek2007}.

Based on the aforementioned issues and empirical findings, 
we first develop a simple model of labor migration and show through agent-based simulation (ABS) 
that the model initiates the aggregation of agents from a random initial distribution. 
Then the obtained spatiotemporal population distribution and economic circumstances 
are compared to empirical observations known as Zipf's and Okun's laws. 
One of the salient features of this study lies in its attempt to integrate 
spatial aggregation phenomena 
and statistical empirical regularities into a single model in the simplest possible way. 
This integration has not been achieved thus far, 
due in part to a lack of theoretical foundations that properly 
characterize the concentration of population from a microscopic perspective. 
Our approach allows for illuminating the multifaceted aspects of urbanization 
comprehensively and coherently. 
We then obtain theoretical explanations for and uncover 
the elements fundamental to the emergence of aggregation phenomena in human society.

The remainder of this paper is organized as follows. 
In Sec.~\ref{sec: framework} we develop a model of urbanization using active Brownian particles 
and delineate the simulation scheme. 
In Sec.~\ref{sec: results} we provide simulation results and derive the corresponding macroscopic model. 
Subsequently, we highlight its relevance to the classical Keller--Segel model 
(a classic model describing the aggregation behavior of biological organisms called taxis) and 
introduce a key concept called \emph{econotaxis} to characterize the reproduced aggregation phenomena. 
Through approximation, we then identify the transition between aggregation and diffusion regimes, 
and compare certain qualitative behaviors derived from this model with empirical data.
Finally, in Sec.~\ref{sec: discussion} we summarize the main 
findings and discuss them both
regarding previous studies and from the perspective of potential relevance to future work.

\section{\label{sec: framework} Methods}

\subsection{\label{subsec: mathematical model} Mathematical model}

We introduce a stochastic model of labor migration, 
following the framework proposed in \cite{Schweitzer1998}. 
Let us consider $N$ identical Brownian agents migrating 
in a domain $\Omega = [0, L]^2\subset \mathbb{R}^2$, 
with $L$ being the system size. 
Brownian agents are characterized by their internal states $\theta_i$ and positions $r_i \in \Omega  \,(i=1, \ldots, N)$. 
Moreover, they interact indirectly with each other via their external environment \cite{Schweitzer2007, Castellano2009}. 
Each agent is assigned either the employed ($\theta_i = 0$) or unemployed ($\theta_i = 1$) status, 
and their spatial movements are governed by either of the following Langevin equations depending on their current status:
\begin{align}
\dot{r_i} &= \sqrt{2D_l} \, \xi_i (t)  \quad (\theta_i = 0), \label{eq: C0} \\
\dot{r_i} &= \sqrt{2D_n} \, \xi_i (t) + F(t) \quad (\theta_i = 1), 
\label{eq: C1}
\end{align}
where $\xi_i(t)$ represents white Gaussian noise with zero mean 
and $\ens{\xi_i(t) \cdot \xi_j(t')} = \delta_{i,j} \delta(t-t')$. 
$D_n$ and $D_l$ are some constants satisfying $D_n \gg D_l$. 
This relation implies that employed agents are almost intrinsically immobile compared to the unemployed 
because people in employment reasonably prefer stability in their lives and are thus less motivated to migrate 
than the unemployed ones, who have incentives to travel even long distances to obtain a job. 
We also suppose that agents transition their status from $\theta_i=0$ to $\theta_i=1$
with a transition probability $k^-$, and vice versa with $k^+$. 

In order to specify $F(t)$, which is crucial in the spatiotemporal dynamics of population distribution, 
we first determine $f(r)$ as follows: 
\begin{align}
f(r) &=\alpha \grad \ln \left[Y\left(l(r)\right) + 1\right],   \label{eq: econotaxis term} \\ 
Y(l) &= A \, l^{\beta},   \label{eq: normal production function}  
\end{align} 
where $f(r)$ is a deterministic force exerted on the agents located at $r$, 
$\alpha$ and $A$ are some constants, 
$l(r)$ is the density of the employed agents, 
$Y(l)$ is a production function, 
and $\beta$ is a positive exponent smaller than $1$ 
\footnote{The condition $\beta < 1$ ensures \emph{decreasing returns to scale}, 
which refers to situations where increasing input by a factor of $\eta$ increases the output 
by a factor \emph{less} than $\eta$ \cite{Fujita2013}}. 
The reason behind this specification of $f$ will be discussed in Sec.~\ref{sec: discussion} 
relative to empirical findings from multiple related fields.

With regard to $F(t)$ in Eq.~(\ref{eq: C1}), 
we consider a spatial discretization of the entire domain $\Omega$ into square boxes 
with a unit length $h$. 
Each of them is denoted by $\Lambda_{mn}$, 
with $m$ and $n$ denoting spatial indices. 
Notice that 
a unique $\Lambda_{mn} \subset \Omega$ exists for any given $r_i(t)$
such that $r_i(t) \in \Lambda_{mn}$. 
We define 
$F(t)$ by
\begin{align}
&F(t)  \nonumber
\\ &=   \label{eq: central difference}
\frac{\alpha}{2h}
\begin{pmatrix}
\ln [Y(l_{m+1,n})+1] - \ln [Y(l_{m-1,n})+1] \\
\ln [Y(l_{m,n+1})+1] - \ln [Y(l_{m,n-1})+1]
\end{pmatrix}
, \\
& l_{m, n}(t) = \frac{1}{h^2} \sum_{j=1}^{N}  
\mathds{1}_{\Lambda_{mn}} (r_j(t)) \delta_{\theta_j(t), 0}, 
 \label{eq: l histogram} 
\end{align}
where $\mathds{1}_{I} (r) = 1 \; (r \in I); \, 0 \; (r \notin I)$ and 
$\delta_{\theta,\theta'}$ is the Kronecker delta. 
Note that $F$ in Eq.~(\ref{eq: central difference}) 
plays the same role as $f$ in Eq.~(\ref{eq: econotaxis term}), 
except that it is reformulated for ABS using the central difference method. 
In practice, $\{l_{m, n}(t)\}_{m,n}$ is equivalent to a uniformly binned histogram 
representing the spatial distribution of employed agents.

\subsection{\label{subsec: simulation} Simulation scheme}

We used the Euler--Maruyama method for the time discretization of Eqs.~(\ref{eq: C0}) and (\ref{eq: C1}) 
with an interval of length $1$. 
The transition between the two statuses (employed and unemployed) 
was also implemented during each time step. 
All agents update their positions and status simultaneously, constituting a unit time step. 
The entire updating process was iterated throughout the simulation. 
Simulations were performed considering periodic boundary conditions.

\begin{figure}[htbp]
\begin{center}
\includegraphics[width=8.6cm]{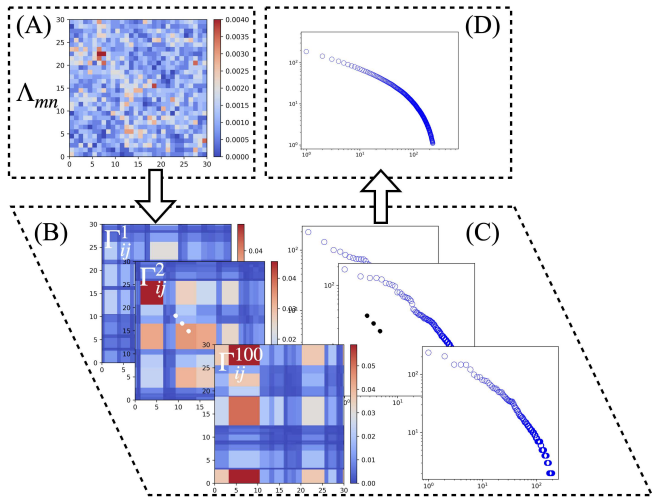}
\caption{Schematic view of computing 
a mean rank-size distribution by applying different arrangements of cities. 
(A) ABS is implemented 
using the fine uniform grid $\{\Lambda_{mn}\}$. 
(B) The entire region is randomly divided into rectangles 100 times 
using coarse, nonuniform grids $\{\Gamma^k_{ij}\}_{i,j}$. 
(C) Rank-size distributions are produced based on the different configurations of cities 
corresponding to the coarse, nonuniform grids $\{\Gamma_{ij}^k\}_{i,j}$. 
(D) A mean rank-size distribution is obtained by averaging over 
all rank-size distributions produced in (C). 
In the following simulations, we prepare 289 cities in the $30 \times 30$ field ($L=30, h=1$) by applying a hundred $17 \times 17$ nonuniform grids generated randomly. 
} 
\label{fig: city arrangement}
\end{center}
\end{figure}

The proposed model was also investigated in terms of two empirical laws. 
First, we clarified the definition of a city in the simulation 
to produce rank-size distributions. 
Perhaps unsurprisingly, 
what defines a city is debatable \cite{Gabaix2004}. 
Here, we adopted a simple approach, where we equated cities with the entire region divided into smaller rectangular areas 
using a \emph{nonuniform} spatial grid. 
More specifically, 
we separated the domain $\Omega$ into smaller rectangles 
by applying a coarse nonuniform spatial grid, represented by $\{\Gamma_{ij}\}$, 
having a fixed number of straight lines on each side 
with their intervals randomly specified. 
After that, we determined each rectangular area defined by that nonuniform grid as an equivalent of a city. 
Furthermore, to mitigate the stochastic effect of the aforementioned arrangement of cities on the resulting rank-size distributions, 
we ``filtered'' the same spatial 
population distribution produced via ABS 
by 100 different nonuniform grids, 
$\{\Gamma^k_{ij}\}_{i,j} (k=1, \ldots, 100)$, 
and obtained a mean rank-size distribution, following the same procedure using these grids 
(see Fig.~\ref{fig: city arrangement}). 
The average size of a city must be effectively larger than the unit spatial length $h$ 
such that the scaling relationship between the migration of agents and the average city size is not violated.
In summary,  the grid
$\{\Gamma^k_{ij}\}_{i,j}$ is nonuniform, generated numerous times, and coarse compared to 
$\{\Lambda_{mn}\}_{m,n}$, which is a fine uniform grid with a unit length $h$ used only for computing our agent-based model.

Regarding Okun's law, we investigate whether fractional changes 
in the total output, $\Delta Y_{tot} / Y_{tot}$, are (anti-)correlated with 
contemporaneous changes in the overall unemployment rate, $\Delta \mu$. 
Note that 
$Y_{tot}(t) = \sum_{m,n} Y\left(l_{m,n}(t)\right)$ 
and $\mu(t) = \left(\sum_{i=1}^N \delta_{\theta_i (t),1}\right)/N$ in the present model. 
We plot the data $(\Delta \mu, \Delta Y_{tot}/Y_{tot})$ for every time step.

\subsection{\label{subsec: previous model} Relation to a previous approach}

In Sec.~\ref{subsec: mathematical model},  
we presented a simple model of labor migration. 
In this section, we delineate the key limitation of the model proposed in \cite{Schweitzer1998},  
which was briefly mentioned in Sec.~\ref{sec: introduction}. 
Quite different from the present model, in that previous study, 
the deterministic force [corresponding to Eq.~(\ref{eq: econotaxis term})]
along with a modified production function [corresponding to Eq.~(\ref{eq: normal production function})] was specified as follows: 
\begin{align}
f(r) &= \grad \omega(r), \quad 
\omega (r) = \frac{\delta Y}{\delta l}  (l(r)),  \nonumber \\
Y(l)  &\propto [A_0 +\exp (a_1 l - a_2 l^2)] \, l^\beta,  \label{eq: nonlinear production function}
\end{align}
where $r$ represents the position, $\omega (r)$ indicates a wage field, $l(r)$ is the density of employed agents, 
$Y(l)$ is the production output as a function of labor, and $A_0, a_1, a_2, \beta$ are positive constants. 
The less compelling aspect of this formulation is its reliance 
on a somewhat artificial modification---a nontrivial 
exponential term in Eq.~(\ref{eq: nonlinear production function})---applied to a standard production function, 
which serves as the basis for the initiated aggregation. 
Consequently, $f$ works on unemployed agents as an attraction force only in a certain middle range of $l$. 
Therefore, aggregation would 
not arise from a random initial distribution of employed agents unless $l$ is initially set high enough. 
Nonetheless, aggregation does occur in \cite{Schweitzer1998} 
because exponential nonlinearity is also assumed in the rates $k^+, k^-$ of transition between two states 
(employed and unemployed) to help $l$ increase and reach that ``special'' range. 
In other words, without these technical conditions, city-like regions would not appear, let alone coexist. 
Thus, we contend that the aforementioned formulation 
does not offer explanations plausible enough for the spatial concentration phenomena observed globally.

\section{\label{sec: results} Results} 

\subsection{\label{subsec: aggregation} Aggregation phenomena}

\begin{figure}[htbp]
\includegraphics[width=\columnwidth]{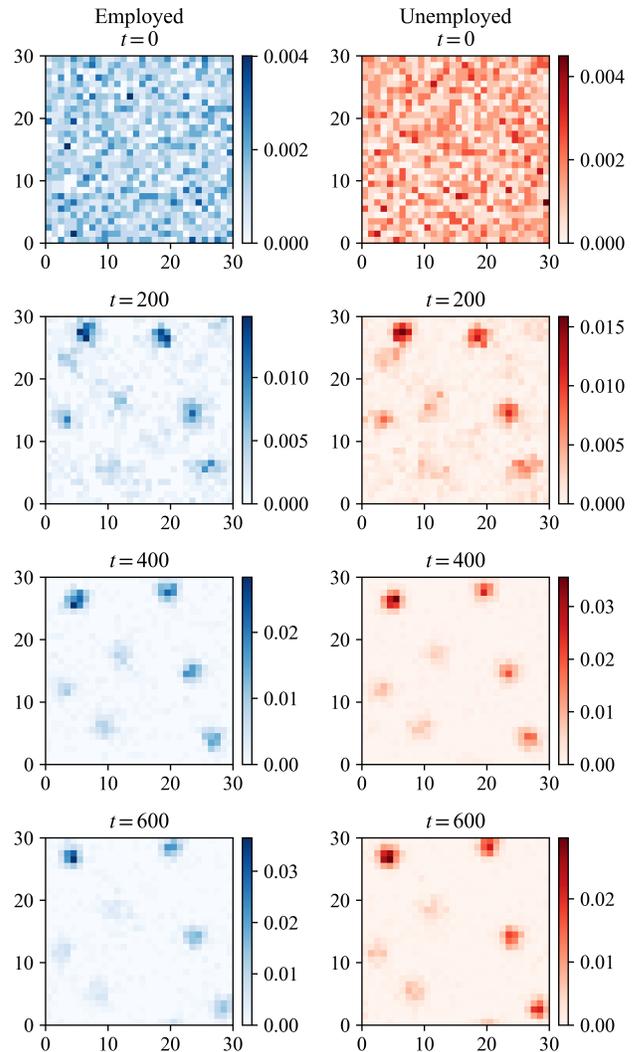}
\caption{Snapshots demonstrating the aggregation of agents produced by ABS. 
The parameters are as follows: $N=4000, D_n=0.3, D_l=0.03, \alpha=1, A=1, \beta=0.67, k^+=0.3, k^-=0.55$. 
This set of parameters applies to the results in all figures that follow unless otherwise noted. 
The colored bars display the fractions of employed (left) and unemployed (right) agents in a uniformly discretized box 
relative to the total number of agents in each respective employment status at the time presented.}
\label{fig: aggregation1}
\end{figure}

Based on the method described in Sec.\ref{subsec: simulation}, 
we implemented ABS to investigate the behavior of our model, 
focusing primarily on the spatiotemporal dynamics of the agent distributions. 
Figure~\ref{fig: aggregation1} shows an evolution of the spatial distributions 
of the employed and unemployed agents, illustrating how they undergo the process of self-organization 
from a spatially disordered to an ordered state.
Distributed randomly at the initial stage, these agents separate into aggregations before $t=200$. 
Later the clusters become tighter, and only some of them 
continue to intensify in magnitude while others disappear (from $t=200$ to $t=400$). 
In the following, we will discuss the mechanism of the observed aggregation phenomena.

In Eq.~(\ref{eq: C1}), two opposing forces compete with one another: \emph{attraction} and \emph{dispersion} forces. 
An attraction force arises from the differences 
in economic circumstances between neighboring places, 
resulting from the differences in the number of people hired at these places 
[see Eqs.~(\ref{eq: central difference}) and (\ref{eq: l histogram})]. 
Meanwhile, a dispersion force is produced by random fluctuations of the agents in motion. 
If attraction overrides dispersion, unemployed agents start forming clusters. 
Once this occurs, the effect amplifies: 
the more unemployed agents present, the more employed agents will exist
because they constantly transition their employment status from unemployed to employed and vice versa. 
This leads to a higher production output, which results in even stronger attraction forces. 
This positive feedback allows for the emergence of disproportionately highly populated areas. 
In Eq.~(\ref{eq: C0}), in contrast, only a dispersion force is present. 
Thus, aggregation of employed agents is caused solely by clustered unemployed agents transitioning their job status.

\begin{figure}[htbp]
\begin{center}
\includegraphics[width=\columnwidth]{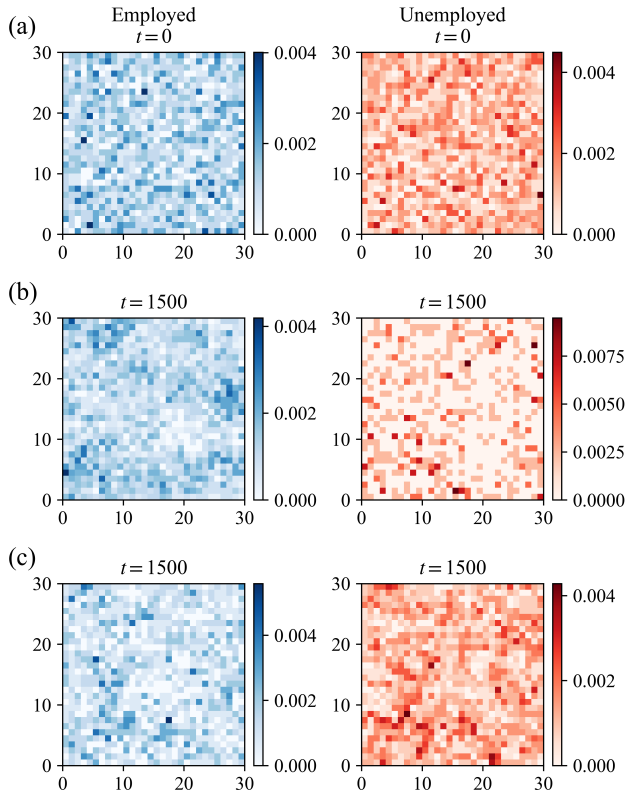}
\caption{Top two rows present snapshots produced by ABS 
where no aggregation occurs owing to a significant imbalance of transition rates. 
The parameters are the same as those in Fig.~\ref{fig: aggregation1}, 
except that the transition rates are chosen as (a) $k^+=0.9, k^-=0.1$ and 
(b) $k^+=0.1, k^-=0.9$. 
The bottom row (c) presents a snapshot demonstrating that aggregation does not occur when $D_n \approx D_l$. 
The parameters are the same as those in Fig.~\ref{fig: aggregation1} 
except that $D_n=D_l=0.3$. 
}
\label{fig: no aggregation} 
\end{center}
\end{figure}

Regarding employment status transitions, the balance between transition probabilities  
can be a critical factor in determining whether agents end up in aggregations. 
According to our simulations, the aggregation of agents is prevented 
when either of the transition probabilities is significantly larger than the other 
[Figs.~\ref{fig: no aggregation}(a) and \ref{fig: no aggregation}(b)]. 
We suggest that this result is because a lopsided pair of transition probabilities produces 
a disproportionately large number of agents of one type 
and a few of the others, as the balance $k^+/k^-$ fundamentally 
determines the overall and local employment rates. 
If most agents are employed, the population distribution flattens out before 
some places have the chance to gain momentum and grow at the expense of others.
If most agents are unemployed, in contrast, 
few employed agents are available to ramp up local production, 
weakening the attraction force sufficiently 
to stop changing the particular orientation in the movements of unemployed agents. 
Therefore, maintaining the ratio significantly far from one 
prevents aggregation in the present model. 
The influence of transition rates on aggregation will be discussed further in Sec.~\ref{subsec: migration}.

Although the condition $D_l \gg D_n$ reflects our primitive assumption on labor mobility, 
as mentioned in Sec.~\ref{subsec: mathematical model}, 
it is noteworthy to see if the relationship is reflected in our model appropriately. 
Figure~\ref{fig: no aggregation}(c) shows in fact that 
aggregation does not occur 
when the magnitudes of intrinsic mobility of employed and unemployed agents are comparable with one another. 
This indicates that the difference in their intrinsic tendency to migration 
contributes to the emergent aggregation behavior, 
which could also be an interesting indicator of urbanization when analyzing empirical data.

 \subsection{\label{subsec: Zipf} Empirical laws} 
 
\begin{figure}[htbp]
\begin{center}
\includegraphics[width=8.6cm]{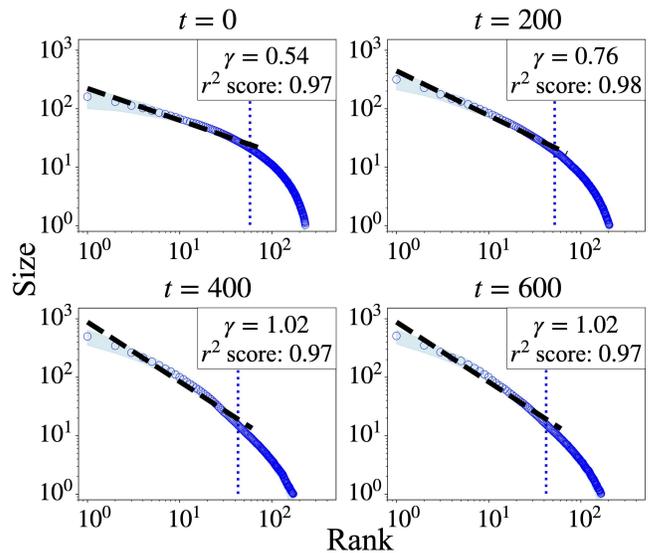}
\caption{Evolution of the mean rank-size distribution for the data 
presented in Fig.~\ref{fig: aggregation1}. 
The data are presented in blue circles. 
The blue-shaded areas represent the standard deviations due to applying 
100 different city configurations. 
The fitting interval 
was truncated at the 75th percentile (i.e., the top $25\%$), as indicated by the dotted vertical line. 
The truncated distribution was regressed linearly (black dashed line) 
using the ordinary least squares method. 
The fitting interval was chosen to be consistent with  
previous empirical analysis performed on U.S. census data \cite{Eeckhout2004}. 
}
\label{fig: Zipf1}
\end{center}
\end{figure}

After verifying that the proposed model reproduces aggregation phenomena, 
we investigate how rank-size distributions evolve over time, as well as 
their relevance to Zipf's law (Fig.~\ref{fig: Zipf1}). 
Note that none of the rectangular areas 
with a population of zero or one are 
considered cities, 
either in the data of Fig.~\ref{fig: Zipf1} or in the following discussion.
First, high-ranking cities (above the $75$th percentile) 
appear to fit well with a straight line 
at all times presented in the figure on a logarithmic scale. 
Notably, as time progresses, 
the slope becomes steeper, and the exponent [i.e., $\gamma$ in Eq.~(\ref{eq: Zipf})] 
gradually approaches $1$ (the \emph{Zipf exponent}) 
without sacrificing the $r^2$ values 
and then stabilizes around 
(from $t=400$ to $t=600$). 
In other words, 
the city size distribution exhibits Zipf's law corresponding to the aggregation.

Furthermore, the standard deviation (owing to how we arrange cities) 
remains small at all ranks and at all times presented.
Although, during an initial stage, the aforementioned method causes moderately large fluctuations 
in the sizes of high-ranking cities, 
these fluctuations gradually disappear as the agents form clusters 
(see Figs.~\ref{fig: aggregation1} and \ref{fig: Zipf1}).  
Empirical studies have demonstrated that the law applies to 
data from various countries despite their varying definitions of an administrative city \cite{Veneri2016}. 
The small fluctuations shown in Fig.~\ref{fig: Zipf1} by the shading 
may suggest a similar type of robustness of the system for different units of analysis.

\begin{figure}[htbp]
\begin{center}
\includegraphics[width=8.6cm]{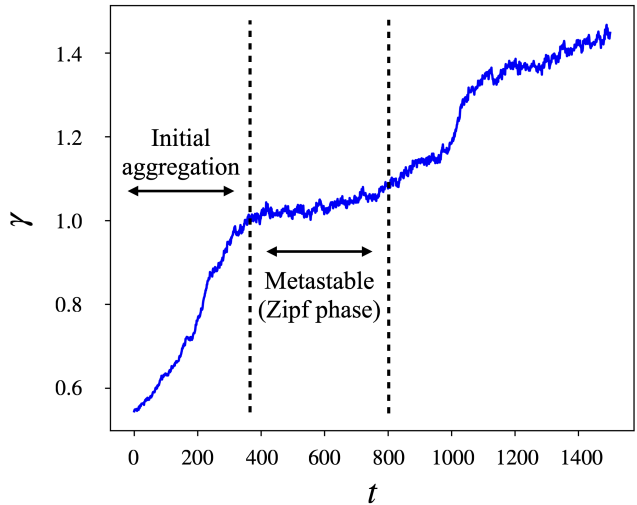}
\caption{Time series of the rank-size exponent in Fig.~\ref{fig: Zipf1} for an extended time window. 
The $r^2$ score is above $0.96$ at all $t$. 
The period when Zipf's exponent appears to be stable (from $t \approx 350$ to $800$) 
suggests that the system is in a metastable state. 
Only fluctuations that are large enough can bring the system out of this state.}
\label{fig: metastable}
\end{center}
\end{figure}

Furthermore, the rank-size distribution seems to stabilize after $t=400$ (Fig.~\ref{fig: Zipf1}).  
To investigate its long-term behavior, we plotted the evolution of the mean rank-size exponent over a longer period 
(Fig.~\ref{fig: metastable}). 
Note that $r^2$ values score more than $0.96$ at any moment. 
We find that in an initial stage (from $t=0$ to $t \approx 350$), the exponent $\gamma$ generally increases monotonically 
as a result of aggregation. 
However, the increase rate exhibits a significant slowdown, followed by a relatively long period (from $t \approx 350$ to $800$)
during which the exponent remains. 
We call it a ``Zipf phase.''

As illustrated in Fig.~\ref{fig: metastable}, however, the exponent starts growing again after the Zipf phase, 
especially with a characteristic sudden increase at $t \approx 1000$. 
During and after the Zipf phase, even established clusters constantly fluctuate 
owing to the random force in Eq.~(\ref{eq: C1}), and smaller clusters potentially merge into nearby larger ones, 
consequently producing a sudden increment in the value of the exponent. 
This highlights the separation of timescales for forming clusters and their possible integration.  
We therefore suggest that Zipf's law does not necessarily represent an equilibrium 
\footnote{This is the most prevailing view on Zipf's law in literature \cite{Gabaix2004, Hsu2012, Arshad2018}}, 
but  rather a \emph{metastable} state where large fluctuations might free the system from being trapped.
Although the exact time and value of the exponent where noticeable slowdowns in the rate of increase occur vary 
occasionally (due to different initial conditions and randomness in the motion of agents), 
such declines and the consequent metastability of the aforementioned phenomenon are typical in most cases.
Our discussion on metastability may provide a clue regarding 
why many urban systems have their rank-size exponents close to 1 
but still demonstrate an overall increasing tendency, as discussed in Sec.~\ref{sec: introduction}.

\begin{figure}[htbp]
\begin{center}
\includegraphics[width=8.6cm]{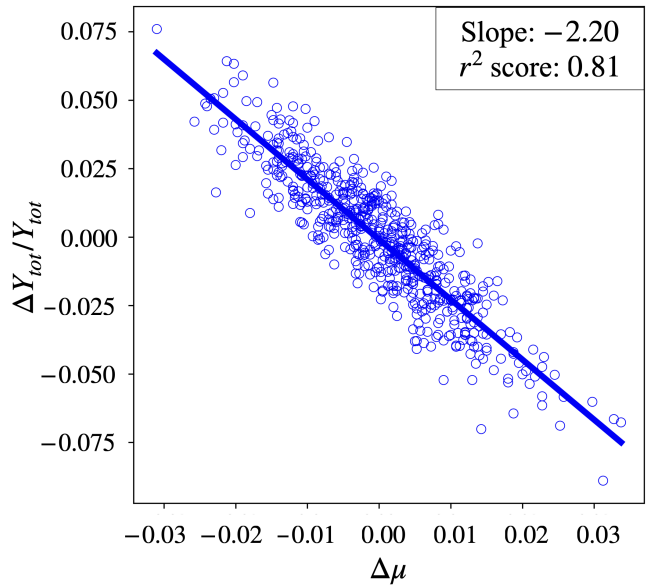}
\caption{
Anticorrelation between changes in the unemployment rate 
and corresponding growth rates in the production output obtained from the same data as in Fig.~\ref{fig: aggregation1} 
(from $t=0$ through $t=600$). 
Linear regression (solid line) corresponding to the plotted data (circles).}
\label{fig: Okun1}
\end{center}
\end{figure}

To understand how labor mobility relates to economic circumstances in the present model, 
we investigated the relationship between 
changes in the unemployment rate and growth rates in the total output (Fig.~\ref{fig: Okun1}). 
In the figure, data are plotted in the $\Delta \mu - \Delta Y_{tot}/Y_{tot}$ plane, 
where each dot represents the relationship between these two variables at each time step. 
The data demonstrate a clear anticorrelation between those two variables. 
In particular, the regressed line has a slope of $-2.20$ with an $r^2$ score of $0.81$. 
For example, the reproduced value of the slope is highly reminiscent of Okun's coefficient in the U.S. economy, 
which is estimated to be around $2$, as mentioned in Sec.~\ref{sec: introduction}. 
The ability of our model to reproduce a coherent coefficient may be indicative of its proper underlying mechanisms.

\subsection{\label{subsec: econotaxis} Econotaxis}

Sec.~\ref{subsec: aggregation} stated that the proposed model reproduces aggregations of agents, 
which is characterized by the orientational movements of unemployed agents 
in response to their potential socio-economic advantages. 
This process can be understood as a positive \emph{taxis} emerging in social contexts. 
In biology literature, taxis is typically referred to as a behavioral response 
where an organism directs its movement in response to an external stimulus \cite{Stevens1997}. 
Mathematically, it can be described by a biased random walk \cite{Grunbaum1998}. 
In the present model, the characteristic movement of unemployed agents 
is particularly initiated by a local economic activity in which employed agents are engaged. 
Therefore, we call this positive feedback process involved in our model ``econotaxis'' 
\footnote{Note that the new concept \emph{econotaxis} is in line with the idea of self-reinforcement in urbanization, 
 which some traditional theories have implied but not necessarily modeled explicitly 
 \cite{Boddy1999, Krugman1991, Krugman1992, Ottaviano2004}}.

Furthermore, we can derive the following Fokker--Planck equations from our microscopic model \cite{Risken1996}: 
\begin{align}
n_t  &= D_n  \, \nabla^2 n - \grad \cdot \left( n f \right)  - k^+ n + k^- l,  \label{eq: nt} \\  
l_t  &= D_l  \, \nabla^2 l + k^+ n - k^- l,  \label{eq: lt}
\end{align}
where the densities of unemployed and employed agents are denoted by $n(r,t)$ and $l(r,t)$, respectively. 
The aforementioned macroscopic system indicates that our model reproduces the clustering of agents similarly to 
 how the classical Keller--Segel system \cite{Keller1970, Horstmann2003} initiates aggregation phenomena called 
\emph{chemotaxis}. 
This is a popular type of taxis by which certain chemicals affect cell migration. 
The Keller--Segel system is composed of two variables: $u, v$, which respectively refer to 
the density of cells and the concentration of chemicals and are given by 
\begin{align}
u_t &= d_u \nabla^2 u -  \grad \cdot \left( u \, \grad \chi (v) \right), \label{eq: ksu}
 \\
v_t &= d_v \nabla^2 v  + a u - b v , \label{eq: ksv}
\end{align}
where $\chi(v)$ is called a sensitivity function determining how migrating cells react to the chemoattractants. 

The relationship of Eqs.~(\ref{eq: ksu}) and (\ref{eq: ksv}) to Eqs.~(\ref{eq: nt}) and (\ref{eq: lt}) in terms of aggregation 
is apparent, which allows us to make at least two further arguments. 
First, the above described investigation offers an interesting perspective on the nature of the labor force in general. 
Comparing econotaxis with chemotaxis, employed and unemployed agents can be identified as
chemoattractants and cells, respectively. 
This suggests that although workers and jobless agents may sound like complete opposites, 
they play complementary roles during the aggregation process: 
Employed agents function as ``attractants,'' helping the unemployed navigate the world 
to enable them to live off better employment opportunities. 
In other words, unemployed agents receive from employed agents a ``signal'' of benefit for their survival 
(i.e., employment opportunities) 
and respond to it while they constantly switch roles. 
Their well-coordinated and intertwined relationship amplifies the initial small heterogeneity and 
allows certain places to develop as highly populated. 
In this manner, the proposed model suggests urban areas may emerge out of nowhere in human society.

Second, a vast amount of analytic investigation has revealed that several (often biological) chemotactic behaviors modeled 
by systems such as Eqs.~(\ref{eq: ksu}) and (\ref{eq: ksv}) can be understood mathematically
as blow-up phenomena \cite{Horstmann2003, Hillen2009, Arumugam2021}.  
This indicates that econotaxis is also related to blow-up phenomena 
in the system of Eqs.~(\ref{eq: nt})~and~(\ref{eq: lt}). 
Its implication will be elaborated on in Sec.\ref{sec: discussion} from a social science perspective.

Although Eqs.~(\ref{eq: nt}) and (\ref{eq: lt}) may fall under the same class as the traditional Keller--Segel system, 
a crucial difference exists between them. 
Contrary to most Keller--Segel models motivated by chemotactic behavior, the proposed system is mass-conserved, i.e., 
$\int_{\Omega} \left[n(r,t) + l(r,t)\right] \, dr = N$, 
which cannot be satisfied by systems comprising cells and chemoattractants, such as Eqs.~(\ref{eq: ksu}) and (\ref{eq: ksv}). 
Mass-conserved (in the above sense) Keller--Segel systems have never been explored before, 
as they rarely arise as biological systems \cite{Arumugam2021}. 
However, the current model offers novel insight into the possibility of a finite-time blow-up 
in a mass-conserved Keller--Segel system, and suggests that it might reflect some characteristic feature of human mobility, 
distinguishing it from that of other living organisms.

It should also be noted that Eqs.~(\ref{eq: nt}) and (\ref{eq: lt}) do not provide a complete picture of our microscopic model.  
Because we expect the macroscopic system to blow up in a finite time, 
it is only supposed to capture the aggregation behavior of the microscopic model in an early stage 
where the first cluster of agents is about to develop. 
Therefore, the metastability observed in Sec.~\ref{subsec: Zipf}, for example, 
is out of range of Eqs.~(\ref{eq: nt}) and (\ref{eq: lt}), 
and it may require a different approach to be described properly.

\subsection{\label{subsec: migration} Transition in migration patterns}

\begin{figure*}[tbp]
\includegraphics[width=\textwidth]{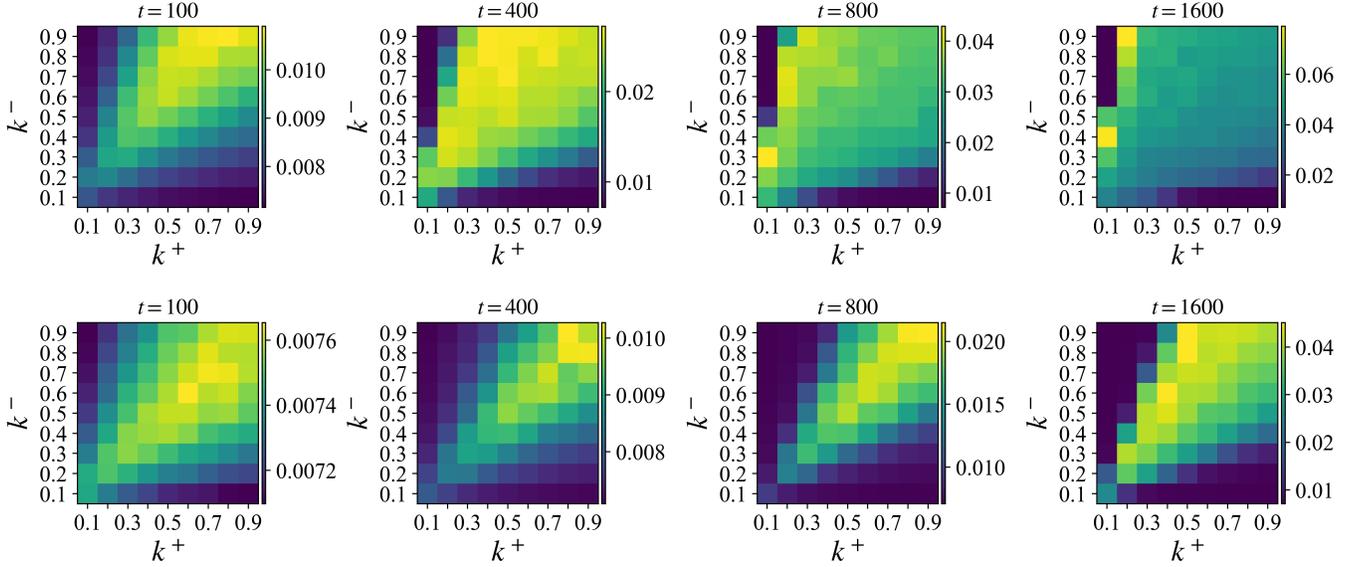}
\caption{
Evolution (from left to right) of $\varphi$ (colored bar) for various values of $k^+$ and $k^-$, 
produced by ABS mentioned in the text. 
The parameters are $A=1$ (top row) or $A=0.6$ (bottom row) with the rest being the same as before. 
The color-coded value $\varphi$ is an indicator for the degree of aggregation, 
and each value across all panels was obtained by averaging the outcomes of ABS over 50 realizations.
}
\label{fig: diagram simulation}
\end{figure*}

\begin{figure*}[tbp]
\includegraphics[width=\textwidth]{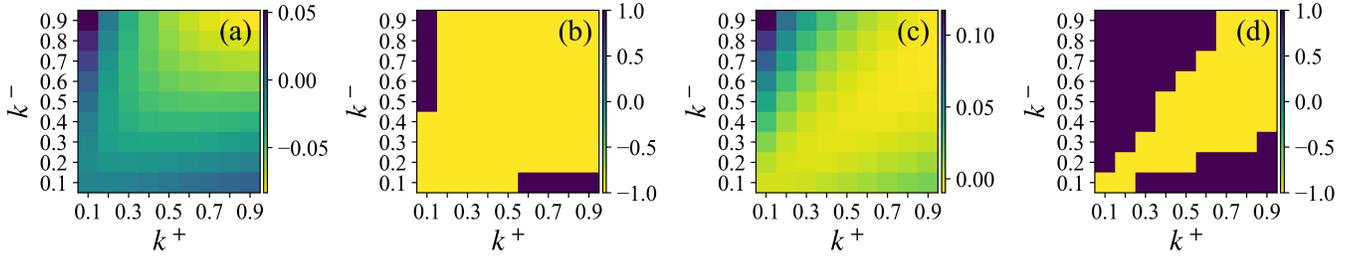}
\caption{
Illustration of $\mathcal{D}(\rho^*)$ [(a), (c)] and $\sgn \left(\mathcal{D}(\rho^*)\right)$ [(b), (d)]
with respect to $k^+$ and $k^-$, 
computed using Eqs.~(\ref{eq: effd}) and (\ref{eq: linitial}). The expression $\sgn (x)$ represents the signum function. 
Other parameters are the same as before, except that $A=1$ [(a), (b)] and $A=0.6$ [(c), (d)]. 
The two-tone figures [(b), (d)] demonstrate the parameter regions for aggregation (yellow) and diffusion (purple), 
while the gradient figures [(a), (c)] represent the varying degrees of initial aggregation with respect to transition rates 
(brighter colors indicate a higher degree of initial aggregation).
}
\label{fig: diagram analytic}
\end{figure*}

In this section, we elaborate on the transition 
between aggregation and diffusion regimes, briefly discussed in Sec.~\ref{subsec: aggregation}, specifically focusing on the transition rates $k^+$ and $k^-$. 
To first gain computational insights into the transition, we implemented extensive ABS for different values of $k^+$ and $k^-$. 
To quantify the degree of aggregation, we introduce the order parameter 
\begin{align}
\varphi = \frac{2}{N(N-1)} \sum_{i < j} \exp \left[-d(r_{i}, r_{j})\right], \label{eq: phi}
\end{align}
where $d(\cdot, \cdot)$ denotes the minimum Euclidean distance under periodic boundary conditions. 
The above parameter is limited to the range $(0,1]$, with higher values indicating more pronounced aggregation among agents. 
Figure~\ref{fig: diagram simulation} demonstrates the time evolution of $\varphi$ with respect to various transition rates 
for $A=1$ (top row) and $A=0.6$ (bottom row). 
In the top row, during earlier times, aggregation seems most facilitated when $k^+ /k^- \approx 1$ 
(Fig.~\ref{fig: diagram simulation}, top left panel). 
However, this tendency shifts later, and the region with the highest degree of aggregation 
moves toward the upper left area where $k^+/k^- \approx 0$ (Fig.~\ref{fig: diagram simulation}, top right panel). 
Dark-colored outer areas along both axes (in all top panels) indicate diffusive behavior. 
The bottom row follows the same trend, although the smaller $A$ consistently delays the onset of aggregation.
In short, the aggregation regime exists around the central area of the parameter space, flanked by the diffusion regime on both sides. 
Additionally, the parameter region with the highest degree of aggregation shifts progressively from the diagonal to the upper left area.

We now provide analytical insights into the identified transition and its time-varying characteristics observed through ABS. 
Following the approach in \cite{Schweitzer1998}, 
one can assume that the reaction process described in Eqs.~(\ref{eq: nt}) and (\ref{eq: lt}) 
relaxes to a quasistationary state quickly, 
i.e., $k^+ n \approx k^- l$. 
This simplifies both equations to 
\begin{align}
\rho_t  \approx \grad \cdot [\mathcal{D}(\rho)  \grad \rho], \label{eq: flux}
\end{align}
where we define $\rho = n + l$ and 
\begin{align}
\mathcal{D} (\rho) = \frac{1}{k^+ + k^-}  \left[D_l k^+ +  \left(D_n - \beta \frac{A l^\beta}{A l^{\beta} + 1}\right) k^-\right]. 
 \label{eq: effd}
\end{align}
Under the quasistationary approximation, we find
\begin{align}
n  \approx \frac{k^-}{k^+ + k^-} \rho, \quad  l \approx \frac{k^+}{k^+ + k^-} \rho. \label{eq: initial}
\end{align}
Using the reduced equation indicated above, 
we first provide an informal qualitative argument on the transition between the two distinct dynamical regimes: 
the agents should aggregate if $\mathcal{D} \left(\rho(r,t)\right) < 0$, 
while they should diffuse if $\mathcal{D} \left(\rho(r,t)\right) > 0$. 
In addition, when the inequality $\mathcal{D} \left(\rho(r,t)\right) < 0$ holds for some $t$ (i.e., aggregation regime), 
the smaller the value of $\mathcal{D}$ is, the more pronounced aggregation should be manifested. 
Therefore, within the aggregation regime, a smaller ratio $k^+/k^-$ should cause the agents to be concentrated 
more extremely in the long term.

We now evaluate $\mathcal{D} \left(\rho(r,t)\right)$ during an early stage. 
Because the agents are initially uniformly distributed across the domain, 
we expect 
\begin{align} 
\rho(r,t) \approx  \rho^*. \label{eq: linitial}
\end{align}
for all $r \in \Omega$ and small $t$, with $\rho^* = N/|\Omega|$. 
Using Eqs.~(\ref{eq: effd})--(\ref{eq: linitial}), 
we can compute $\mathcal{D} (\rho^*)$ for various $k^+$ and $k^-$. 
Figure~\ref{fig: diagram analytic} presents the results 
for [Fig.~\ref{fig: diagram analytic}(a)] $A=1$ and [Fig.~\ref{fig: diagram analytic}(c)] $A=0.6$, 
along with their corresponding sign patterns [Figs.~\ref{fig: diagram analytic}(b) and \ref{fig: diagram analytic}(d)], 
illustrating the predicted parameter regions for the aggregation regime. 
Distinctly, $\mathcal{D}(\rho^*)$ decreases as the ratio $k^+/k^-$ approaches $1$ 
[Figs.~\ref{fig: diagram analytic}(a) and \ref{fig: diagram analytic}(c)], 
aligning with our previous speculation on the initial behavior of the system (Sec.~\ref{subsec: aggregation}).
The figures indicate both aggregation and diffusion regimes, with the former dominating the central region surrounded by the latter in outer areas.

Let us finalize this section by comparing the ABS results with the theoretical argument provided above, 
accompanied by brief explanations. 
First, the predicted initial aggregation behavior is in line with the observed patterns in the ABS 
[see the leftmost panels of Fig.~\ref{fig: diagram simulation} 
compared to Figs.~\ref{fig: diagram analytic}(a) and \ref{fig: diagram analytic}(c)]; 
aggregation is most facilitated along the diagonal area where $k^+/k^- \approx 1$. 
The first step toward understanding the mechanism underlying this observation is 
to note, from Eq.~(\ref{eq: effd}), that $\mathcal{D}$ is a decreasing function of $l$ and for a sufficiently large $l$, the term $Al^\beta /(Al^\beta + 1)$ approximates $1$. 
Hence, the functional form of $\mathcal{D}$ requires $D_n < \beta$ for aggregation to occur. 
More importantly, Eqs.~(\ref{eq: effd}) and (\ref{eq: initial}) also require $k^+/k^-$ not to be either too small or too large 
as this would mean either $l$ would be too small, resulting in $\mathcal{D} >0$, 
or the term $D_l k^+$ would be too large relative to $(\beta - D_n) k^-$, also resulting in $\mathcal{D} >0$.  
Second, once the condition $\mathcal{D} <0$ holds, Eq.~(\ref{eq: effd}) guarantees that 
the magnitude of $k^-$ relative to $k^+$ determines the strength of the concentration of agents. 
In fact, the dynamic shift from the diagonal to the left-hand peripheral area with the highest degree of aggregation, 
as documented in Fig.~\ref{fig: diagram simulation}, 
is consistent with our analytical prediction that a smaller ratio $k^+/k^-$ eventually 
reinforces the aggregation process to a greater extent 
[see the rightmost panels of Fig.~\ref{fig: diagram simulation} 
compared to Figs.~\ref{fig: diagram analytic}(b) and \ref{fig: diagram analytic}(c), 
especially with an emphasis on the leftmost outer edges of the bright central regions]. 
Additionally, the observed delay in the onset of aggregation with the smaller $A$ (noted by comparing the bottom to the top row in Fig.~\ref{fig: diagram simulation}) is easy to predict from the form of Eq.~(\ref{eq: effd}). 
It is worth noting that some quantitative deviations exist between our analytical and simulation results, 
such as near the critical region around the upper left areas. 
Therefore, a thorough investigation of Eqs.~(\ref{eq: nt}) and (\ref{eq: lt}) may be helpful 
to fully uncover the transition between the aggregation and diffusion regimes.

\subsection{\label{subsec: empirical data} Empirical data analysis}

\begin{figure}[htbp]
\begin{center}
\includegraphics[width=8.6cm]{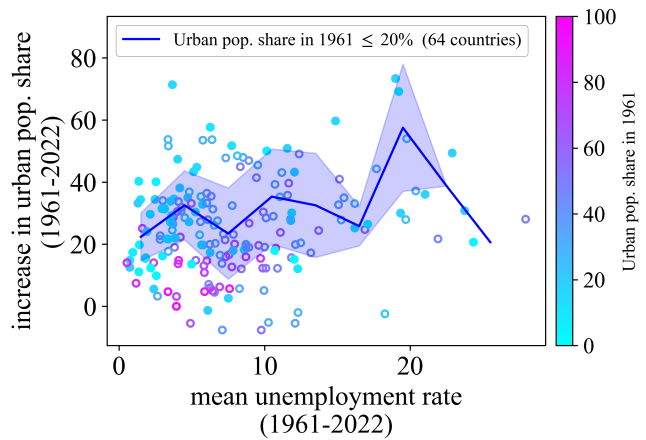}
\caption{
National-level relationship (blue solid line) between the nationwide mean unemployment rate 
and the increase in the share of urban population from 1961 to 2022, 
derived from World Bank Open Data \cite{worldbank}. 
Each colored circle represents data from a single country (185 countries in total). 
For a precise assessment of our theoretical argument,  
only data from countries with less than a 20\% urban population share in 1961 (64 countries: filled circles) 
were used to establish this relationship.  
The blue shading represents the standard deviation for the line, 
and the colored bar illustrates the initial share of the urban population in 1961. 
}
\label{fig: data}
\end{center}
\end{figure}

We now seek to empirically examine our theoretical prediction that  
a higher unemployment rate is associated with a higher degree of aggregation in the long term. 
We analyzed data (colored circles) from 185 countries obtained from World Bank Open Data \cite{worldbank}. 
Countries are not included in the figure if and only if they lack data 
on either the national unemployment rate or the urban population. 
Figure~\ref{fig: data} illustrates the observed relationship between the mean unemployment rate 
and the increase in the share of the urban population at the national level. 
For a precise assessment of our theoretical prediction, we used data solely from countries that had been ``under-urbanized,'' 
which we defined as those whose urban population share was below 20\% in 1961, 
and computed the mean increase (blue line) in the urban population share over the extensive time period 
from 1961 to 2022 for a given mean unemployment rate. 
More precisely, the mean increase in the share of urban population for a given unemployment rate was obtained 
by binning the data from the 64 countries into nine bins on the horizontal axis, with each bin having a width of 3. 
The means within those bins were then calculated. 
For a comprehensive understanding, however, all the available data (185 countries) are shown in the figure. 
We also note some characteristic features of the time series data. 
The unemployment rate fluctuates but generally remains stable over the time period, 
with occasional worldwide drops during major catastrophic economic events. 
The mean standard deviation in the unemployment rate over the years is 1.64. 
In contrast, the share of the urban population tends to increase continuously over time in most countries. 

Let us next interpret the data considering our theoretical argument. 
The mean increase (blue solid line in Fig.~\ref{fig: data}) in the percentage of urban population 
as a function of the mean unemployment rate demonstrates a gradual increase for small unemployment rates. 
It then appears to reach its peak at an approximately 20\% unemployment rate, beyond which it decreases quite abruptly. 
The first segment of this behavior seems to align quite well with our qualitative prediction 
that the unemployment rate is associated with the extent of long-term aggregation. 
In contrast, the latter part of the behavior is hard to interpret 
primarily because the amount of data available for the highest mean unemployment rates 
(especially at more than 25\%) is insufficient. 
While the model predicts a discontinuous transition to the diffusion regime at some high unemployment rate, 
it is currently not clear whether it corresponds to the observed drop in the data. 
Overall, we conclude that the empirical data, particularly the initial phase of the behavior, 
appears to support our theory qualitatively. 
We note, however, that the observed peak could be due to outliers, 
as the number of available data points generally decreases with the mean unemployment rate. 
Therefore, the results should be taken with caution.

\section{\label{sec: discussion} Discussion and concluding remarks}

\begin{table*}[htbp] 
    \caption{Relevant concepts and laws.}
    \label{tab: concepts}
    \begin{tabularx}{2\columnwidth}{l*{1}{Y}}
	\hline \hline
	Concept or law & Brief explanation  \\
	\hline  
	Zipf's law (rank-size rule) & Power-law relationship between rank and city size with an exponent of 1  \\
	Gibrat's law & Proportionate city growth with constant mean and variance  \\
	Okun's law & Anticorrelation between output growth and increase in unemployment rate  \\ 
	Cobb--Douglas production function & Simple formula connecting labor force and output \\
	Fechner's law & Human sensation grows in proportion to the logarithm of stimulus intensity \\ 
	Active Brownian particles & Brownian particles moving autonomously \\
	Taxis (e.g., chemotaxis) & Orientational movement of living organisms in response to external stimuli  \\
	\hline \hline
    \end{tabularx}
\end{table*}

In this section, we first discuss the motivation behind the attraction force specified in the present model. 
Unlike \cite{Schweitzer1998}, we assumed that unemployed agents prefer \emph{economically active} places. 
In this context, we imply that unemployed agents tend to migrate toward areas where a large economic output is produced. 
In other words, unemployed agents are drawn to places where, for example, well-paid or secure jobs are available, 
and their chances of being provided with those opportunities improve as the local economic production increases. 
Although living in those places can be accompanied, in reality, by various issues such as higher living costs and crime rates, 
we postulate that they offer significant benefits in almost every aspect of one's life, 
and the benefits of dwelling in urban areas generally outweigh the disadvantages. 
Empirical evidence provides insights supporting this argument \cite{Berry2008, Boddy1999}. 
Studies have shown that economically disadvantaged people are attracted to cities, 
urbanward migration is a response to growing economic opportunity, and 
a positive relationship exists between wages and city size.
In economics, production output can be computed as a function of the number of workers. 
The simplest and the most commonly used form is the Cobb--Douglas production function \cite{Douglas1976}, where the 
output is assumed to increase proportionally to the number of workers raised to some power. 
In our model, all of the above discussions are 
summarized in Eqs.~(\ref{eq: econotaxis term}) and (\ref{eq: normal production function}). 
The logarithm in Eq.~(\ref{eq: econotaxis term}) is based on Fechner's law, originally proposed in psychophysics, 
which states that sensation grows proportional to the logarithm of stimulus intensity \cite{Nutter2010}. 
Specifically, we regard production output as a perceivable stimulus to which 
unemployed agents respond in search of better employment opportunities. 
The logarithmic representation and encoding of stimuli in the brain have been confirmed extensively for nonphysical stimuli, 
such as numerical digits \cite{Dehaene2003, Nieder2003} and the quality of user experience in telecommunications \cite{Reichl2010}.  
In the present case, the production output could alternatively be perceived in the form of information through ads or rumors, 
which at least partially supports the use of Fechner's law. 
Note that we assume less highly nonlinear functions in 
Eqs.~(\ref{eq: econotaxis term}) and (\ref{eq: normal production function}) 
and constant transition rates $k^+, k^-$, as opposed to \cite{Schweitzer1998}, 
making in-depth mathematical analysis far more accessible.

Our simple approach to urbanization might cast doubt on a major economic theory, which states that 
increasing returns to scale [i.e., $\beta > 1$ in Eq.~(\ref{eq: normal production function})] 
is a prerequisite for urban agglomeration \cite{Fujita1996, Duranton2004}. 
Instead, the present model 
suggests that a straightforward concave production function can cause the concentration of the population 
(note, however, that most economic theories center on economic activities 
and may not necessarily be comparable with the present model). 
In addition, the results in Sec.~\ref{subsec: Zipf} imply that the two empirically-observed 
regularities may not arise from some specific details of social or economic interactions. 
Instead, they might arise from the mechanism underlying urbanization, which has rarely been explored previously. 
In particular, regarding Zipf's law, Simon's model \cite{Simon1955} has been criticized 
for slowing down and taking an infinite amount of time 
to converge to Zipf's law as one seeks to obtain a power-law exponent close to $1$ \cite{Krugman1996, Gabaix1999}. 
We resolved this limitation by developing a model that demonstrates a coherent power-law-like behavior in finite time 
as we predict a finite lifetime of the macroscopic system 
and the consequent transition to a metastable state of the microscopic model. 
The relation to Girat's law, 
however, is not clear  
because the present model exhibits migration-based---i.e., spatially correlated---city growth, 
in contrast to Gibrat's law, which assumes proportionate city growth with constant mean and variance.

From a sociological perspective, the suggested finite-time blow-up, as discussed in Sec.~\ref{subsec: econotaxis}, 
may provide a potential answer to a long-standing question in urban science: 
is there a saturation point in urbanization (i.e., an ``urban maturity'') \cite{Tisdale1942, Berry2008}? 
The social science literature states that whether the process reaches equilibrium is still unknown \cite{Tisdale1942, Davis1955}, 
even though some scholars are in favor of the idea of saturation \cite{Berry2008, Fujita2013}. 
Our numerical results considering a new concept of econotaxis suggest otherwise: 
urbanization can be a relentless process.

In summary, 
we have presented a simple microscopic model of urbanization to reveal the underlying fundamental mechanism. 
Despite its simplicity, the model initiated the aggregation of agents and reproduced some statistical regularities 
in human societies and economies. 
Subsequently, we proposed a novel concept called econotaxis 
to characterize the self-assembly of the labor force. 
Moreover, we revealed the complementary roles played by employed and unemployed agents 
during the process of aggregation. 
Our model offers a unified theoretical approach to the study of urbanization 
and provides insight into the possibility that the macroscopic aggregation phenomena and 
Zipf's and Okun's laws arise from the same underlying mechanism, 
which has not yet been adequately discussed. 
By laying the proper microscopic foundations, 
this study integrates multiple crucial aspects of urbanization 
while staying mathematically tractable. 
The resulting aggregation model [Eqs.~(\ref{eq: nt}) and (\ref{eq: lt})]
is \emph{per se} of particular interest because it is mass-conserved 
in a way that most systems describing the migration of organisms, including chemotaxis, are not. 
By employing the quasistationary approximation, 
we identified the transition between aggregation and diffusion regimes both analytically and computationally. 
The analytical argument that we presented captures the transition qualitatively but fails to offer a quantitative understanding of this transition. 
Finally, we compared the model with extensive real-world data 
on labor statistics and urban indicators and found moderate consistency between them. 
The concepts and laws relevant to this paper are summarized in Table~\ref{tab: concepts}.

In the future, 
the present model may be modified to reproduce some of the unique features of urban agglomeration, 
such as the U.S. manufacturing belt or Europe's hot banana \cite{Krugman2011}, 
which may be achieved 
by introducing geographical heterogeneity or spatial constraints. 
Other possible ways to extend the proposed model include incorporating ``congestion effects'' \cite{Berry2008, Boddy1999} 
(i.e., dispersion increases as a place becomes crowded), 
considering an in- and outflux of population or intrinsic population growth and explicitly considering multiple industries. 
However, 
a possible limitation is that the continuum models $(\ref{eq: nt})$~and~$(\ref{eq: lt})$ 
are currently not likely to reproduce Zipf's and Okun's laws because blow-up phenomena are expected to occur. 
Obtaining meaningful analytical insights into these models 
may require further investigation or proper modifications to the models.

The data analyzed in Sec.~\ref{subsec: empirical data} might be subject to several potential limitations. 
First, there is currently no consensus on the definition of an ``urban area'' across nations, which 
could possibly lead to variations in the calculation of urban populations. 
Second, the existing data lack a measure for additional urbanization within urbanized areas, 
which might result in an underestimation of the ongoing intraregional urbanization process in certain areas.
Third, population growth might occur at varying rates in rural and urban areas, 
possibly complicating the direct analysis of rural-to-urban migration within the available data. 
Further studies can explore various indicators for a more multifaceted understanding 
of the underlying mechanism of urbanization.

Finally, as a potential avenue for further research, 
the aggregation phenomena discussed here can be approached in terms of ergodicity, 
which may be achieved by utilizing the ergodicity breaking parameter \cite{Shi2023, Cherstvy2015}.  
Several previous studies indeed have shown that human mobility patterns may be linked to anomalous diffusion processes 
\cite{Brockmann2006, Song2010}. 
Finally, we hope this study contributes to uncovering 
the structures and organizations of our society that allow for its proper functioning and coordination, 
encouraging both theoretical and empirical research on social dynamics.

\section{\label{sec: acknowledgement} Acknowledgements}

I would like to thank Prof.~Kota Ikeda (Meiji University, Japan) for 
insightful discussions and his comments on the manuscript, and 
Dr.~Ryu Fujiwara (Meiji University) and 
Prof.~Hiraku Nishimori (Meiji Institute for Advanced Study of Mathematical Sciences, Japan) 
for their helpful comments on the study.

%

\end{document}